\documentclass[traditabstract]{aa} 
\usepackage{graphicx}
\usepackage{txfonts}
\newcommand{\JMP}{Mart{\'\i}n--Pintado}

\newcommand{\MWC}{MWC\,349}

\newcommand{\Lsun}{L$_\odot$}
\newcommand{\Msun}{M$_\odot$}
\newcommand{\Mdot}{\.M}

\newcommand{\ccm}{cm$^{-3}$}
\newcommand{\dgr}{{\hbox {$^\circ$}}}
\newcommand{\km}{km\,s$^{-1}$}
\newcommand{\dex}[2]{$#1\,10^{#2}$}	
\newcommand{\mkm}{\rm km\,s^{-1}}

\def\ltsima{$\; \buildrel < \over \sim \;$}
\def\simlt{\lower.5ex\hbox{\ltsima}}
\def\gtsima{$\; \buildrel > \over \sim \;$}
\def\simgt{\lower.5ex\hbox{\gtsima}}
\newcommand{\Ha}[1]{H$#1\alpha $}   		
\newcommand{\Hb}[1]{H$#1\beta $}   		
\newcommand{\Hg}[1]{H$#1\gamma $}   		
\newcommand{\Hd}[1]{H$#1\delta $}   		
\newcommand{\He}[1]{H$#1\epsilon $}   		

\newcommand{\pfeiler}{\rule{0em}{3ex}}  		

\newlength{\eempty}

\newcommand{\be}{\begin{enumerate}}
\newcommand{\ee}{\end{enumerate}}
\newcommand{\bi}{\begin{itemize}}
\newcommand{\ei}{\end{itemize}}
\newcommand{\beq}{\begin{equation}}
\newcommand{\eeq}{\end{equation}}

\newcommand{\1}{{\it (i)}}
\newcommand{\2}{{\it (ii)}}

\def\vol#1{{\bf {#1}{\rm,}\ }}
\def\apj   {{\it Ap.~J.~{\rm,}\ }}
\def\apjl  {{\it Ap.~J.~(Letters){\rm,}\ }}
\def\apjs  {{\it Ap.~J.~Suppl.{\rm,}\ }}

\def\aj    {{\it A.~J.{\rm,}\ }}
\def\aap   {{\it Astr.~Ap.{\rm,}\ }}
\def\aas   {{\it Astr.~Ap.~Suppl.~Ser.{\rm,}\ }}
\def\anrev {{\it Ann.~Rev.~Astr.~Ap.{\rm,}\ }}

\def\mnras {{\it M.N.R.A.S.{\rm,}\ }}

\def\rmex  {{\it Rev.\ Mexicana Astr.~Ap.{\rm,}\ }}

\newcommand{\MdotVexp}{\Mdot/\Vexp}
\newcommand{\Vexp}{$\upsilon_{exp}$}
\newcommand{\Lk}{LkH$\mathbf{\alpha}\,101$}
\newcommand{\kms}{\mbox{km~s$^{-1}$}}
\newcommand{\BR}{B\'aez--Rubio}
\begin{document}
%
   \title{\Lk\ at Millimeter Wavelengths}
   \author{C. Thum
          \inst{1}
          \and
           R. Neri\inst{2}
          \and 
           A. B\'aez--Rubio\inst{3}
          \and
           M. Krips\inst{2}
          }

   \institute{
              Instituto de Radio Astronom{\'\i}a Milim\'etrica,
              Avenida Divina Pastora 7, N\'ucleo Central, 
              18012 Granada, Spain\\
              \email{thum@iram.es}
         \and
              Institut de Radio Astronomie Millim\'etrique, 
              300 rue de la Piscine, Dom. Univ. de Grenoble,
              38406 Saint Martin d'H\`eres, France\\
              \email{neri@iram.fr, krips@iram.fr}
         \and
              Centro de Astrobiolog{\'\i}a (CSIC/INTA)
              Ctra de Torrej\'on a Ajalvir, km 4, 
              28850 Torrej\'on de Ardoz, Madrid Spain\\
              \email{baezra@cab.intas-csic.es}
             }
   \date{Received February 29, 2012; }

 
  \abstract{
            We present new millimeter observations of the ionized wind
            from the massive young
            stellar object \Lk, made with the IRAM interferometer
            and 30m telescope. Several recombination lines, including
            higher order transitions, were detected for the first time
            at radio wavelengths in this source. \\
            From three
            $\alpha$--transitions we derive an accurate value for the
            stellar velocity and, for the first time, an unambiguous
            expansion velocity of the wind which is 55\,\km, much slower
            than reported previously, and the mass loss rate is
            \dex{1.8}{-6} \Msun\,yr$^{-1}$.  The wideband continuum
            spectra and the 
            interferometer visibilities show that the density of the wind 
            falls off more steeply than compatible with
            constant--velocity expansion. We argue that these
            properties indicate that the wind is launched from a
            radially narrow region of the circumstellar disk, and we
            propose that slow speed and a steep density gradient are
            characteristic properties of the evolutionary phase where
            young stars of intermediate and high mass clear away the gaseous
            component of their accretion disks. \\
            The recombination lines are emitted close to local thermal
            equilibrium, but the
            higher order transitions appear systematically broader and
            weaker than expected,
            probably due to impact broadening.\\
            Finally, we show that \Lk\ shares many properties with
            \MWC, the only other stellar wind source where radio
            recombination lines have been detected, some of them
            masing. We argue that \Lk\ evades masing at millimeter
            wavelengths because of the disk's smaller size and
            unfavorable orientation.  Some amplification may however
            be detectable at shorter wavelengths. 
           }

   \typeout{end of title page}
   \maketitle
%

%


\section{Introduction}

Radio recombination lines have proved to be a valuable tool for
investigations of the dust enshrouded environments of young massive stars.
These stars are sufficiently 
hot to ionize part or all of the surrounding gas which then emits
recombination lines as a function of its electron temperature, density,
and emission measure. At centimeter wavelengths where recombination
lines typically probe electron densities in the range of $n_e$ =
\dex{}{3} - \dex{}{5} \ccm, the kinematics and other physical
properties of numerous compact and ultracompact HII--regions were
investigated (Roelfsema \& Goss \cite{RRLreview}; Sewilo et
al. \cite{Sewilo};  and references therein).  

Regions of higher electron density, like the inner parts of stellar
winds or the coronae of circumstellar disks, however do often not emit
centimeter recombination lines. The free--free emission from such
regions, with $n_e$ 
$\simgt$\dex{}{6}\,\ccm\ and typical sizes $\sim$\dex{}{-3} pc, are opaque at 
centimeter wavelengths. Sensitive millimeter observations are needed
for such sources, of which the IR-- and radio--bright stellar wind
source \MWC\ may well be regarded as prototypical. The properties of
its inner ionized disk/wind system has been the subject of numerous
studies (B\'aez--Rubio et al. \cite{Moreli}, and references therein). 

In this paper, we report the detection, using both IRAM telescopes,
of several millimeter recombination lines from another radio star
\Lk\ (IRAS\,04269+3510), whose
radio emission is a factor 3 weaker than that of \MWC. The spectral
energy distribution of \Lk\ has long been known to be rising toward
shorter wavelengths (Cohen \cite{Cohen}) which indicates the presence of an
ionized stellar wind. Interferometric observations in the near
infrared ($\lambda\lambda\ 2 - 12 \mu$m) reveal the presence of a hot
circumstellar disk seen nearly face--on (Tuthill et
al. \cite{Tut}). This complex circumstellar environment was already
hinted at in earlier optical/IR spectroscopic investigations (Herbig
\cite{Herbig}; Thompson et al. \cite{Thompson}; Simon \& Cassar
\cite{SC}; Hamann \& Persson \cite{HP}) whose detections of permitted
and forbidden lines demonstrated the presence of high density ionised and
neutral regions as well as low density regions. Velocity--resolved
spectroscopy (e.g. Simon \& Cassar \cite{SC}) furthermore showed some
high density tracers to be 
double--peaked as expected from a rotating disk. Low density tracers
showed the kinematic signature of a slowly flowing ionized wind.

\Lk\ illuminates NGC1579, the brightest patch in the L1482
nebulosity. The star also created a small HII--region, and it must therefore be
hot. The star's spectral type and luminosity class cannot however be
obtained from the optical spectrum, since it does not have any
photospheric absorption lines (Herbig \cite{Herbig}). Instead its
spectral type, near B0.5, is derived from optical photometry, and 
broad band IR photometry (Harvey et al. \cite{Harvey}; Barsony et
al. \cite{Barsony}) gives its luminosity of 2 -- \dex{3}{4} \Lsun.
Such a star at the zero--age main--sequence has an effective
temperature of $\sim$30000 K and a mass of $\sim$15 \Msun\ (Panagia
\cite{Nino}).   

Most physical parameters of \Lk\ depend strongly on the star's distance 
which was somewhat controversial in the past. Estimates ranged from
160 pc placing the star near the Taurus dark cloud (which however is not known
for its massive star formation) up to 800 pc, the value originally
given by Herbig (\cite{Herbig}) on the basis of two optically
identified B--stars. In a recent optical investigation of the
\Lk\ environment, Herbig et al. (\cite{HAD}) found 5 additional
B--stars which each illuminate a small patch of L1482 and are
therefore at a very similar distance as the nebulosity and \Lk. These
authors derive a  distance of the nebulosity of 700 pc which is
the value we use here. Data taken from the literature have been scaled
to this value if necessary.


\section{Observations}
\label{s:obs}

\label{ss:bure} 

\subsection{Plateau de Bure}

The observations were carried out with the IRAM interferometer in its
six-element configuration (March 13, 2010) under conditions
of excellent atmospheric seeing ($\sim$0.23$''$) and low zenith
opacity ($\sim$0.03) at 100 GHz. The wideband correlator WIDEX was
configured to cover the 97.7 -- 101.3\,GHz radio frequency band in both
polarizations with a velocity resolution of 6.0\,\kms. This 3.6 GHz
wide band includes five recombination lines, \Ha{40} and 4 higher
order transitions whose relative
calibration is better than 3\% 
due to their inclusion in the same correlator band.
The narrowband correlator was adjusted to
observe the line transitions in horizontal polarization with a
velocity resolution of 0.24\,\kms. Line frequencies and gaussian fits
are listed in (Tab.~\ref{t:gauss}). 

The radio--frequency bandpass
calibration was performed on 2\,MHz channels using
B0415+379. The slope of the spectrum (Fig.~\ref{f:widex}) was found to be 
within \Lk\ spectral index uncertainties over the 10 to 200 
GHz range. Amplitude and phase were calibrated by interleaving
observations of B0411+341 and B0415+379 every 24\,min. 
To reduce phase noise, visibilities were self--calibrated in phase 
on all baselines using as input the clean--component model.
The flux density scale, which was derived from observations of 3C273 and
adjusted to fit the radio spectrum of MWC349 (Krips et al. \cite{Krips}),
is estimated to be accurate to within 10\%. The synthesized beam of
$1.43''\times  0.99''$                        
was calculated adopting natural weighting. The data were
calibrated and analyzed with the GILDAS \footnote{
http://www.iram.fr/IRAMFR/GILDAS}
software package.

\subsection{30m telescope}

The observations at the IRAM 30m telescope were made during 16--17
October 2011 using the facility receiver EMIR (Carter et al. \cite{Matt})
in the setup where the 3 and 1.3mm bands, each in dual polarization,
were observed simultaneously. Both bands of 4 GHz width were
connected to the autocorreletor WILMA which provides a spectral
resolution of 2 MHz. The 3mm band is centered at 106.7 GHz and
contains the recombination lines \Ha{39} and \Hb{49}. The 1.3mm band is
centered at 257.0 GHz and contains \Ha{29} and \Hg{41}.
Line frequencies are listed in Tab.~\ref{t:gauss} together with
gaussian fits.

Observing conditions were good, with zenith opacities not exceeding
0.15 at 257 GHz. System temperatures ranged between 200 and 250 K at 257
GHz, and were around 100 K in the 3mm band. Line and continuum antenna
temperatures were derived from the usual hot/cold load calibration and
then converted to flux densities using the Jy/K ratios of 6.1 (8.4)
at 107 (257) GHz as given on the Observatory's web
pages. Tab.~\ref{t:gauss} gives statistical errors. The absolute
calibration uncertainty is estimated to be about 15\% at 1\,mm and 10\%
for the longer wavelengths.


\section{Results}

\subsection{Continuum emission}
\label{ss:cont}

\subsubsection{Interferometer data.  }
The $uv$ coverage obtained during the 3.6 hours on--source was 
sufficient to obtain a high quality continuum map
of \Lk. 
The source is slightly offset from the phase center and appears to 
be extended. From fitting the
visibilities in the $uv$--plane we derive a total flux of $230\pm 25$
mJy (Tab.~\ref{t:cont}).  No other source is seen in the primary
beam ($46''$ at 100 GHz)  down to a 
$3\sigma$ limit of 7.2\,mJy. The companion detected by Tuthill et
al. (\cite{Tut}) in the near infrared at $\sim0.17''$ from \Lk\ cannot
be discriminated against the main source due to insufficient angular
resolution. The ``necklace'' of radio sources discovered by Becker
\& White (\cite{BW}) lies outside of our primary field of view.

\begin{figure}
\centering
\includegraphics[width=9cm,angle=0]{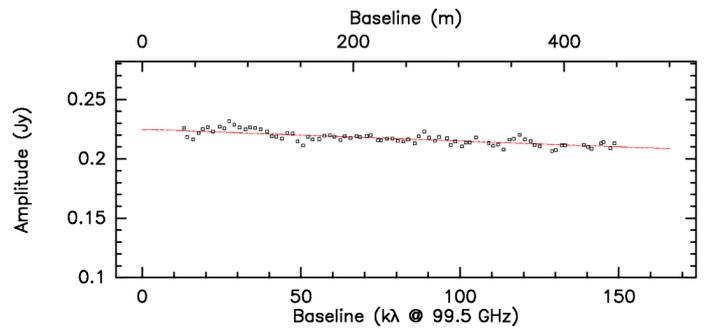} 
\caption{Visibilities of \Lk\ at 99.5 GHz observed with the IRAM
  interferometer over baselines ranging up to 450\,m. The straight
  line represents the fit of a wind model as described in the text. 
  \label{f:cont}
}
\end{figure}

\begin{table}
\caption{Parameters of the continuum source.}      
\label{t:cont}   
\centering                          
\begin{tabular}{lcl}\hline
position & $\alpha$ = 04$^{\rm h}$ 30$^{\rm m}$ 14{\fs}45 & J2000 \pfeiler \\
         & $\delta$ = +35$^\circ$ 16$'$ 24{\farcs}0 &  \\[0.5ex]
flux density &  $230 \pm 25$ mJy & integrated over source \\[0.5ex]
radius $r_1$   &  $0\farcs020 \pm 0\farcs002 $  & where $\tau$(3mm) = 1\\
exponent $\gamma$ & $-2.5 \pm 0.5$
                  & density $\propto (\frac{r}{r_1})^{-\gamma}$\\
$n_0 r_0^2$       & \dex{8.8}{35}\,cm$^{-1}$ & see text\\[0.5ex]\hline
\end{tabular} 
\end{table}

\begin{figure*}
\centering
\includegraphics[width=0.99\textwidth,angle=0]{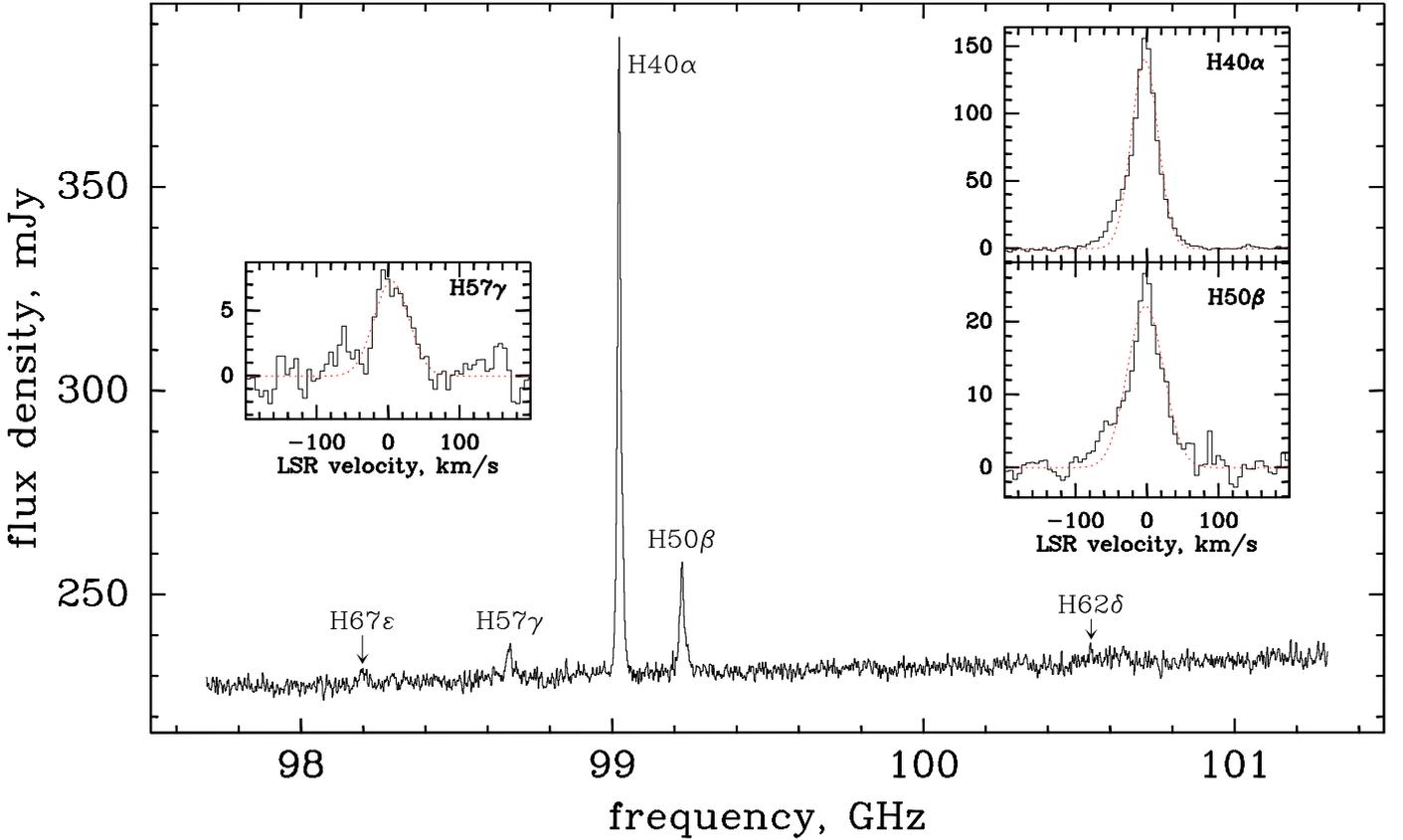}  
\caption{Interferometer spectrum obtained with the new wideband
  correlator (WIDEX) centered at 99.5 GHz. The five recombination lines
  which fall in the 3.6 GHz wide spectrum are labelled. The
  zoomed--in spectra of the 3 clearly detected lines are shown in the
  insets on velocity scales with the continuum subtracted.
  The dotted blue curves indicate gaussian fits whose parameters are
  listed in Tab.~\ref{t:gauss}.
  \label{f:widex}
 }
\end{figure*}
%

The shape of the visibility function (Fig.~\ref{f:cont}), azimuthally
averaged and plotted against $uv$--radius, clearly shows that the 
source is resolved. The visibility amplitude drops by $\sim\!10$\% over
the range of baselines observed. We have tried to fit this data with several
shape functions, including gaussians, but we obtain a satisfactory fit  
only in the case where the electron density $n_e$ falls off with
radius $r$ as a power law ($\frac{r}{r_1})^{-\gamma}$  where the
radius $r_1$ and the exponent $\gamma$ are given in
Tab.~\ref{t:cont}. Density laws of this type are typical of ionized
winds as discussed below. 

The fit assumes 
spherical symmetry of the wind and derives an inner cut off radius $r_0$, 
which turns out to be close to the radius $r_1$
where the source becomes  
optically thick ($\tau = 1$ at 3mm). With the electron density $n_0$ at
$r_0$, the fit gives the parameter $n_0 r_0^2$
which characterizes the strength of the
wind.  The exponent $\gamma$ of the density law 
and the electron temperature $T_e$ are  not well
constrained by the fit. While $T_e\sim\!9500$~K, $\gamma$ ranges
 from $-2$ for the constant--velocity
wind to $-3$ where the wind would be strongly accelerated.  The
derived value of $r_1$ implies that the wind has a small optically
thick core whose blackbody emission amounts to $\sim85$ mJy. 
The  major part
(65\%) of the observed 3mm flux must therefore come from the optically
thin envelope.

\subsubsection{The SED.  }
\begin{figure}
\centering
\includegraphics[width=0.47\textwidth,angle=0]{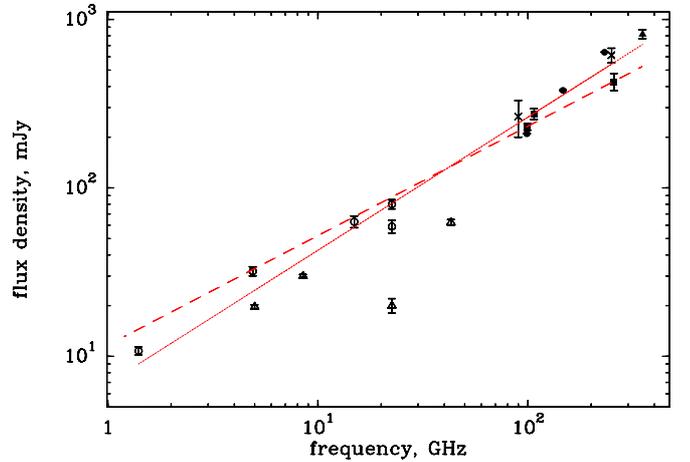}
\caption{Spectral energy distribution of \Lk\ in the radio/millimeter
  regime. Data are taken from Becker \& White \cite{BW} 
  ($\circ$); Gibb \& Hoare \cite{GH}
  ($\bigtriangleup$), Altenhoff et al. \cite{ATW} ($\times$), Sandell
  et al. \cite{Sandell} ($\blacktriangle$), priv. comm. by M. Krips 
  ($\bullet$), and this paper  ($\blacksquare$).
  The continuous red line is a weighted fit to the
  observations and has a slope of $\alpha = 0.86$ in
  logarithmic units. The dashed red line is a fit to the same data
  where the slope has been fixed to its canonical value $\alpha = 0.60$.  
  \label{f:sed}
 }
\end{figure}

We combined the flux density obtained at 99.5 GHz from the
interferometer with the observations at 106 and 257 GHz from the 30m
telescope (Tab.~\ref{t:gauss}) and with data from the literature to
construct the spectral energy distribution (SED) of \Lk\ (Fig.~\ref{f:sed}).
Since the source is embedded in an extended HII--region (Becker \&
White \cite{BW}), we use at the lower frequencies only observations of high
angular resolution. At frequencies above $\sim30$ GHz, the HII--region
is much weaker than \Lk\ in beams smaller than $30''$. From the 5\,GHz
map shown in Fig.~2 of Becker \& White (\cite{BW}) we estimate that the
total flux of the HII--region is less than 30 mJy. At millimeter
wavelengths the HII--region is even weaker, and its contribution in
the 3mm beam of the 30m telescope is less than 10 mJy and completely
negligible in the interferometer beam and even in the 1mm beam of the
30m telescope.

The source has the inverted spectrum typical of an ionized wind as has
already been reported extensively in the literature (Simon \& Cassar
\cite{SC}; Becker \& White \cite{BW}; Hamann \& Persson \cite{HP};
Altenhoff, Thum, and Wendker \cite{ATW}; Gibb \& Hoare \cite{GH};
Sandell et al. \cite{Sandell}). While the
spectral slope $\alpha$ (defined as  $S_\nu \propto \nu^{\,\alpha}$) is 0.6 for
the canonical case of an isothermal plasma expanding isotropically at constant
velocity (Panagia \& Felli \cite{PF}), we derive the slightly higher
value of $\alpha = 0.86\pm0.03$ from the data shown in Fig.~\ref{f:sed}. We
have excluded from 
our fit the VLA--A observations at 22.5 and 43 GHz which appear anomalously
low, possibly due to calibration problems. However, if this data
  were retained,  the result ($\alpha = 0.89\pm0.07$) would be
  essentially the same.

The higher than canonical value of $\alpha$ is strongly
supported by the \1 the slope of the 3.6 GHz wide interferometer
spectrum and \2 the fit to the visibilities. From the slope of the 
spectral baseline (Fig.~\ref{f:widex}) we derive $\alpha=
0.90\pm0.10$, fully compatible with the value
derived from the SED. The power law exponent $\gamma$ derived from the
fit of the visibilities is directly related to the spectral slope of the SED
(Panagia \& Felli \cite{PF}; Olnon \cite{Olnon}), and we derive 
$\alpha = 0.95\pm 0.15$, again higher than  the canonical value. 

%

The data used for representing the SED were taken by many independent projects
over the course of two  decades. Any time variations of the source or
calibration biases may
then distort the derived $\alpha$. We therefore give higher weight to the
values of $\alpha$  derived from the interferometer spectrum and
visibilities, and adopt a best estimate of $\alpha = 0.90\pm0.10$. 
As a corollary, the electron density of the ionized wind falls off 
like $r^{-2.4}$, slightly steeper than in the canonical case.

\begin{table*}
\caption{Recombination lines (rest frequency $\nu$) observed and
  gaussian fits. $S_L$ is
  the peak line flux density, $\upsilon_{LSR}$ the LSR velocity of
  the center of the line, and $w$ its full width at half
  power. $S_C$ is the continuum flux density. }      
\label{t:gauss}   
\centering                          
\begin{tabular}{l r r@{\,}c@{\,}l r@{\,}c@{\,}l r@{\,}c@{\,}l 
                    r@{\,}c@{\,}l r@{\,}c@{\,}l r@{\,}c@{\,}l l} \hline
transition &$\nu$ [GHz] \pfeiler
           &\multicolumn{3}{c}{$S_C$ [mJy] $^a$} 
           &\multicolumn{3}{c}{$S_L$ [mJy]} 
           &\multicolumn{3}{c}{$\upsilon_{LSR}$ [\km]} 
           &\multicolumn{3}{c}{$w$ [\km]} 
           &\multicolumn{3}{c}{$\int S_L d\upsilon\ ^b$)} 
           &\multicolumn{3}{c}{$L/C$ [\km] $^c$)} 
           & remark \\[1ex]\hline
\Ha{40} & 99.022958  & 230 &$\pm$& 3 & 141 &$\pm$&3& -2.6&$\pm$&0.2
        & 43.8 &$\pm$& 0.3 & 7.10&$\pm$&0.1&30.9&$\pm$&1.2 
        & non--gaussian wings\pfeiler\\
\Hb{50} & 99.225214  & 231 &$\pm$ & 3  & 22 &$\pm$ &3& -1.5&$\pm$ &0.8
        & 60.0 &$\pm$&2.2  &1.56 &$\pm$&0.1 & \\
\Hg{57} & 98.671897  & 229 &$\pm$ & 3  &7.4 &$\pm$ &0.7&5.2&$\pm$ &2.3
        & 64.1 &$\pm$&6.5  &0.54&$\pm$&0.05 & \\
\Hd{62} & 100.539631 & 233 &$\pm$ &3   & \multicolumn{3}{c}{$< 3.0$} \\
\He{67} & 98.199388  & 228 &$\pm$ &3   & \multicolumn{3}{c}{\simlt2.5}\\[1ex]
\Ha{39} & 106.737363 & 275 &$\pm$&18 &226 &$\pm$& 8& 0.4 &$\pm$& 0.4
        & 32.8&$\pm$ &1.2  & 7.80&$\pm$&0.28 & 28.4&$\pm$&2.5 
        & non--gaussian wings \\
\Hb{49} & 105.301864 & 274 &$\pm$     & 18 &38 &$\pm$ & 10&$-2.0$&$\pm$& 2.0 
        & 32.9& $\pm$&5.0  &1.35& $\pm$ &0.20 &  \\[1ex]
\Ha{29} & 256.302051 &428  &$\pm$&34 &1250&$\pm$&34 &$-1.6$ &$\pm$&0.2
        & 30.3&$\pm$ &0.5  &42.3&$\pm$  &0.6 & 98.8 & $\pm$& 7.0
        & non--gaussian wings \\
\Hg{41} & 257.635490 &430 &$\pm$ & 34   &\multicolumn{3}{c}{$< 85$} & & & &
        & & & & & & \\[1ex]
\hline                                   
\end{tabular}
\begin{list}{}{}
\item[$^a$)] 30m values are derived from the  spectroscopic baseline.
\item[$^b$)] derived from numerical integration over the line profile,
             in units of [Jy\,\km]
\item[$^c$)] velocity--integrated line--to--continuum ratio
\end{list}
\end{table*}

\subsubsection{Wind structure. }

From the continuum data we derive some physical properties of the
ionized wind. Following
Panagia \& Felli (\cite{PF}) and adapting their equations to our
non--canonical slope of $\alpha = 0.90$, we get 
\begin{equation}
   \dot{M} / \upsilon_{exp} = 3.2\,10^{-8} \ {\rm M}_\odot\ {\rm
     yr}^{-1} ({\rm km\,s}^{-1})^{-1}  
   \label{e:MV}
\end{equation}
where $\dot{M}$ is the mass loss rate and $\upsilon_{exp}$ the
expansion velocity of the wind. 
Assuming that the wind is isotropic, we calculate the characteristic
radius $R(\nu)$, inside of which half of the emission at frequency
$\nu$ is generated, and the electron density $n_e$ at this radius as
\begin{eqnarray}
   R(\nu) &= &11.1\ \left[\frac{\nu}{300\ {\rm GHz}}\right]^{-0.55} 
                                                          \label{e:Rnu}\\
   n_e &=& 5\,10^7\ \left[\frac{R(\nu)}{10\,{\rm a.u.}}\right]^{-2.4} 
                                                          \label{e:ne}
\end{eqnarray}
where $R(\nu$) is in astronomical units, a.u.,  and $n_e$ in cm$^{-3}$.

For the frequency of 99.5 GHz at which our interferometer observations
were made, eq.~\ref{e:Rnu} predicts a characteristic radius of 20
a.u. or 29 mas. This is somewhat larger than $r_1$, the radius where
the source becomes optically thick at 3mm (Tab.~\ref{t:cont}). This
relation,  $R(\nu) \simgt r_1$, must be expected given the different
definition of these radii and the fact that most of the emission 
comes from the optically thin envelope.  

Encouraged by this agreement, we proceed to use these parameterizations 
to infer several important properties of the ionized wind. First, 
we obtain an estimate of the electron density,
$\sim3\,10^7$ \ccm, at 13 a.u. Emission around 260 GHz, the frequency
of our \Ha{29} and \Hg{41} transitions, comes
predominantly from this region.  Correspondingly, our various
transitions near 100 GHz come from $R(\nu)\sim\!26$ a.u. where the
electron density is of the order $8\,10^6$ \ccm. These densities are
several orders of magnitude higher than those in the well studied compact
HII--regions (Gordon \& Walmsley \cite{GW}). Unusual behavior of the
millimeter recombination lines as indeed found here
(sect.~\ref{ss:ratios})  may then be expected. 

Secondly, the emission at 300 GHz, the highest frequency where the
non-canonical slope is known to hold, originates at radii near 11
a.u. (eq. \ref{e:Rnu}). This is inside the dusty torus 
which surrounds the star at $\sim15$ a.u. (Tuthill et al. \cite{Tut}).
Either the wind is launched from this inner dust--free region or, more
likely in our view, the assumption of an isotropic stellar wind breaks down at
radii smaller than the dust disk. 

As a consistency check of the derived parameterization of the stellar
wind, we calculate 
the beam--averaged continuum brightness temperature  measured with the
interferometer. Using the relation between flux density $S_\nu$ (in
mJy) and brightness temperature $T_{mb}$ (in K) for a gaussian beam of
equivalent FWHP $\theta_b$ (in arcsec), valid at $\lambda = 3$\,mm 
$$ T_{mb} = 0.122\ S_\nu\ \theta_b^{\,-2} $$
we derive a beam--averaged brightness temperature of 17.5 K.  
We estimate the beam dilution adopting the source radius predicted by
eq.~\ref{e:Rnu} to be a factor of  1.7\,$10^{-3}$.
The  source brightness temperature then
follows as $\sim\!10^4$ K, typical of photo--ionized nebulae.
Given the approximatins involved, the derived brightness temperature
is also in reasonable
agreement with the electron temperature 
derived by Becker \& White (\cite{BW}) who obtain 7000~K from directly
fitting the fall--off of the visibility function with $uv$--radius.
This agreement confirms that the ionized wind is optically thick even at
millimeter wavelengths.

\subsection{\mbox{The \boldmath $\alpha$}--transitions}
\label{ss:alpha}

We have detected in \Lk\ 6 recombination lines, among which there are 3
$\alpha$-transitions, in the 1 -- 3 mm 
wavelength range. The only other stellar wind source where
radio recombination lines have so far been detected is \MWC\ (Altenhoff et
al. \cite{ASW}) where the lines are found masing over a wide range of
wavelengths (\JMP\ et al. \cite{MBTW}; Thum et al. \cite{laser}).
It is therefore worth checking whether the recombination lines
in \Lk\ are in local thermal equilibrium.

\subsubsection{Line--to--continuum ratios}
\label{sss:L/C}

The line--to--continuum, L/C, ratios offer the best quantitative tool
for investigating any 
strong departures from local thermal equilibrium (LTE). The
velocity--integrated L/C--ratio observed for \Ha{40} in \Lk\ is
$$  \int\frac{S_L}{S_C} d\upsilon = 30.9 \pm 1.2\ \mkm$$
using the values given in Tab.~\ref{t:gauss}.  The L/C--ratio expected for an
optically thin plasma is in the range from 47.6 to 30.6 \km for  LTE
electron temperatures  $T_e^*$ ranging from 7000 to $10^4$ K. 
(Mezger \& H\"oglund \cite{MH}). We note that we have neglected in this
calculation any ionized helium, in line with the low upper limit observed
(sect.~\ref{sss:He}). 

This range of expected L/C--ratios barely includes the observed value. 
We know however that the L/C--ratio is
reduced in wind sources compared to the fully optically thin
case. Altenhoff et al. (\cite{ASW}) 
derive a reduction factor of $\frac{2}{3}$ for the canonical case of a
constant--velocity wind. This reduction factor is further
decreased to 0.53 (Rodr{\'\i}guez et al. \cite{RZH}) if the electron
density falls off like $r^{-2.4}$, as  suggested above.  
We then derive LTE electron temperatures of 7400~K for the canonical
$r^{-2}$ wind and 6300~K for the $r^{-2.4}$ case.
These temperatures bracket the 7000~K derived at the end of
Sect.~\ref{ss:cont} and obtained by Becker \& White (\cite{BW}).
This agreement supports  our view that the $\alpha$--transitions are
emitted near LTE. 

This view is further strengthened by the L/C--ratios which we derive
for \Ha{39} and \Ha{29} (Tab.~\ref{t:gauss}). Both ratios are fully
compatible with our 
\Ha{40} value which is the most precise of the 3
measurements. Extrapolating the mean of the 3mm $L/C$--ratios to 256
GHz like $\nu^{1.1}$,  as appropriate for plasma under LTE
conditions, we predict a \Ha{29} $L/C$--ratio of 90 \km, which
agrees with the observed value within  observational error.

\subsubsection{Systemic velocity }
\label{sss:vel}

The center velocities of our three $\alpha$--transitions are in the range of 
$-2.5$ to $+0.4$ \km. The individual values are mildly
incompatible with each 
other, probably due to small departures from gaussianity of the line
profiles. A weighted average gives the systemic velocity of the wind
in the local standard of rest, LSR, as
$$ \upsilon_{sys} = -1.5 \pm 0.3\ \mkm$$
which we also take to be the radial velocity of the star
(Rodr{\'\i}guez \cite{Luis}). Very similar
velocities were obtained from Br$\alpha$ ($-2\pm2$ \km) 
and Br$\gamma$ ($-1\pm2$ \km) by Simon \& Cassar (\cite{SC}). 
The star appears to be at rest in its natal molecular cloud as demonstrated
by the very close agreement in velocity ($-1.7$ \km) of its optically
thin HCN emission   (Pirogov \cite{Pirogov}). Velocities
of Balmer transitions, by contrast, appear to be shifted to the blue
by $-2 \ldots -4$ \km\ (Herbig et al. \cite{HAD}). Other optical lines
also appear to be blue shifted by similar amounts, as found by the same
authors. This behavior is to be expected if dust in the disk obscures
part of the receding flow.  Velocities of absorption lines are
slightly blue shifted as well by $\simlt1$\,\km\ as shown by the
vibration ally excited 
transition of CO at $\lambda4.7\,\mu$m (Mitchell et
al. \cite{Mitchell}), the 1--0 rotational transition of CO (Knapp et
al. \cite{Knapp}; Gibb et al. \cite{Gibb2010}), and atomic hydrogen
(Dewdney \& Roger 
\cite{DR}). These neutral components may evidence a slow expansion
of the molecular cloud in which \Lk\ is embedded.

\subsubsection{Wind expansion velocity and rate of mass loss}
\label{sss:exp}

We estimate the wind expansion velocity \Vexp\ from \Ha{40} which is
the line with the best observed profile. Due to the low inclination of
the (nearly face--on) disk, its contribution to the line profile
appears at low radial 
velocities near the line center, whereas the wind dominates the line wings.
This decomposition of the line profile into disk and wind components
is also justified on grounds of the symmetry properties of the profile. Whereas
the gaussian line core is rather symmetric, the wings are not. Again
due to the near face--on geometry, emission from the approaching and
receding parts of the wind must be expected to differ as a
consequence of optical depth and stimulation. The blue--shifted part
of the wind should then be brighter, as is indeed observed (right
inset of Fig.~\ref{f:widex}).  We detect emission from the wind 
out to $+50$ and $-70$ \km. Taking into account the
thermal broadening of $\sim20$\,\km, we conclude that \Vexp\ is of the order
of 55 \km with an estimated uncertainty of 10 \km. 
We postpone a more detailed discussion of the observed line profiles
to a follow--up paper (B\'aez--Rubio et al., in preparation).

Adopting our estimate of \Vexp = 55 \km\  and using the ratio
\MdotVexp\ derived above (eq.~\ref{e:MV}) we obtain the rate of  mass
loss  of the wind 

\begin{equation}
  \dot{M} = 1.8\,10^{-6}\ {\rm M}_\odot\ {\rm yr}^{-1}    \label{e:Mdot}
\end{equation}

The first derivation of \Mdot\ in \Lk\ (Cohen, Bieging, and Schwartz
\cite{CBS}) combined the radio flux with an optical \Vexp\ based on 
H$\alpha$. The width of  H$\alpha$, $\sim\!\!\!1300$\,\km, was however later
shown to be due to electron scattering (Hamann \& Persson
\cite{HP}), and the derived \Mdot\ was thus much too high. In an
attempt to correct for this effect, Hamann \& Persson model the
H$\alpha$ profile assuming a location, density and temperature
structure of the electron scattering layer. They conclude that the H$\alpha$
line wings outside $\sim\!\!350$\,\km\ can all be explained by electron
scattering. Later determinations of \Mdot\ (e.g. Gibb \& Hoare
\cite{GH}) then used this value as the expansion velocity of the
wind, despite the fact that such large velocities are not supported by
the widths of other optical/NIR hydrogen or helium lines. Their widths
are much smaller (30\ldots60 \km) and more in line with our measurement. 
Since our millimeter observations are not affected by the
uncertainties of dust obscuration and of the precise H$\alpha$ profile
modelling, we think that our measurement
provides the first reliable determination of \Mdot.

\subsubsection{Helium}
\label{sss:He}
The recombination lines of singly ionized helium are offset from the
transitions of hydrogen with the same quantum numbers by $-122$ \km.  
We have inspected the four spectra of $\alpha$--transitions  (see,
e.g. Fig.~\ref{f:widex}) 
and do not find any evidence for ionized helium. We give an upper
limit for its abundance (by number) of
$$ y^+ < 0.03 $$
This low He$^+$ abundance compounds earlier conclusions by Simon \&
Cassar (\cite{SC}) who find that $y^+$ cannot be larger than 0.5 based
on transitions in the near IR. The low $y^+$ is generally attributed to the low
effective temperature of the central star whose spectral type B0.5 
(Cohen \cite{Cohen}) implies a very low luminosity of photons
energetic enough to ionize helium (Panagia \cite{Nino}).

\subsection{Higher order recombination lines}
\label{ss:ratios}

\begin{figure}
\centering
\includegraphics[width=0.47\textwidth,angle=0]{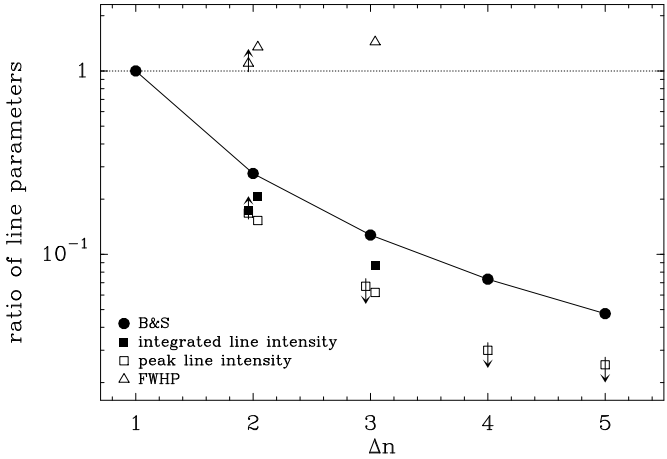}
\caption{Ratios of close frequency pairs. Each ratio relates a
  parameter of a recombination line of order $\Delta n$ ( $=
  n_{upper}-n_{lower} = 1,2,3,4$ for the $\alpha, \beta, \gamma,$ or $\delta$
  transitions) to that of the $\alpha$
  transition at a nearby frequency (see Tab.~\ref{t:gauss}). Parameter
  ratios are shown for the peak line intensity and, whenever
  available, the line full width at half power (FWHP) and the integrated
  intensity.
  The continuous line labelled ``B\&S'' gives the ratios expected for 
  optically thin plasma at low density (Brocklehurst \& Seaton \cite{BS}). 
  \label{f:ratios}
 }
\end{figure}

Higher order recombination lines, i.e. transitions where the principal
quantum number changes by $\Delta n > 1$, are another tool for investigating
departures from thermal equilibrium. Traditionally, the ratio of a
higher order line and an $\alpha$--line at a nearby frequency is used in
order to minimize calibration problems.  
The practicality of this tool at millimeter wavelengths has been
demonstrated on a number of compact HII--regions  by Gordon and
Walmsley (\cite{GW}). The tool was also applied to the stellar wind
source \MWC\ (Thum et al. \cite{beta}), where the line ratios clearly
show that some $\alpha$--transitions are amplified. Here we apply this
tool for the first time to the supposedly  ``ordinary'' stellar wind
source \Lk.

Several close frequency pairs have been selected for this
investigation (Tab.~\ref{t:gauss}). We show their observed
ratios, of both their integrated line flux and the peak line flux,  as
a function of $\Delta n$ in Fig.~\ref{f:ratios}. The thick continuous
line gives the ratio expected for optically thin, low density plasma.
We also show the ratio of the widths of the line pairs. 

We find that all {\it flux} ratios are significantly below their classical
values, with the discrepancy increasing with $\Delta n$,
while their {\it width} ratios are significantly higher than unity and also
increasing with $\Delta n$. We note that close frequency pairs observed in
different wavelength bands and with different telescopes consistently
follow this same  trend.  This behavior is the signature of impact
broadening. 

Motivated by this finding and the inadequacy of gaussians for
fitting the stronger lines (Fig.~\ref{f:widex}), we have fitted the
four transitions with the best observed line profiles with Voigt
functions (Fig.~\ref{f:voigt}). These fits are clearly better than
those obtained with gaussians, particularly near the line centers. 
There are still significant discrepancies, mainly on the blue line
wings, which are discussed below. Tab.~\ref{t:voigt} gives the
parameters obtained from the voigtian fits. Statistical uncertainties
are of the order of $1\ldots2$ \km.  The Doppler widths of all well detected
lines come out to be close to $\sigma = 10$\,\km, corresponding to a 
FWHP $\sim\!24$ \km, a value larger than the thermal line width
of hydrogen at 7000~K (17.9 \km). The resulting Lorentz impact parameter
$\delta$ clearly
increases with $\Delta$n. For the $\alpha$--lines, $\delta$ is smaller
than the Doppler width and impact broadening is barely noticeable. But
already for \Hb{50}, we have $\delta > \sigma$, and the width of the
$\gamma$--line is fully dominated by impact broadening. 

Recombination theory gives the magnitude of impact broadening as
a function of the electron density $n_e$ and the quantum number $n$.
For the range of $n$ of interest here 
$\delta$ is approximated (Walmsley \cite{Malcolm}) as:
\begin{equation}
\delta = 4.2 \left( \frac{n}{100}\right) ^{4.6} n_e $$
\label{e:delta}
\end{equation}
Taking $n_e$ in the various line emitting
regions from our scaling relations eqs. \ref{e:Rnu} and \ref{e:ne},  
we estimate the model $\delta$ as listed in Tab.~\ref{t:voigt}.
Given the approximative nature of these relations, the agreement of
the model $\delta$ with the $\delta$ obtained from the Voigt function
must be considered satisfactory. This agreement supports our 
claim that the higher order transitions are affected or even
dominated by impact broadening. 

This finding may then also explain why the ratio of integrated line
fluxes in close frequency pairs (Fig.~\ref{f:ratios}) is lower than
expected from recombination theory. Impact broadening, while not
affecting the integrated flux of a line, transports a fraction
of the line flux into the wings where it quickly gets lost in the noise.
The losses become more severe with increasing $\Delta n$.
Observations with higher signal--to--noise than those presented here
are needed to confirm this conjecture.

\begin{table}
\caption{Parameters of Voigt function fits and prediction of the
  impact broadening parameter $\delta$.}      
\label{t:voigt}   
\centering    
\begin{tabular}{l@{}ccccl}\hline
     & \multicolumn{3}{c}{fit} &model & \pfeiler \\ \cline{2-4}
\makebox[5em][l]{line} & \ $\sigma\, ^1$ & $\delta\,^2$ & FWHP\,$^3$& $\delta$ &remark\pfeiler
\\[1ex]\hline
\Ha{39} & 10 & 7 & 31 & 3 \pfeiler\\
\Hb{49} &    &   &    & 9 &data too noisy for fit\\[0.5ex] 
\Ha{40} & 10 & 7 & 31 & 3 \\ 
\Hb{50} & 10 &15 & 43 & 10\\ 
\Hg{57} & 10 &20 & 51 & 18\\ 
\Hd{62} &    &   &    & 29 &not detected\\ 
\He{67} &    &   &    & 42 &not detected\\[0.5ex]
\Ha{29} & 10 & 4 & 29 & 2\\ 
\Hg{41} &    &   &    & 7 & not detected\\[0.4ex]\hline
\end{tabular} 
\begin{list}{}{}
\item[$^1$] Gaussian scatter, \km\ (FWHP = $2.35 \sigma$) 
\item[$^2$] impact broadening parameter, \km\ (FWHP = $2\delta$)
\item[$^3$] full width at half power, \km, of the fitted Voigt function
\end{list}
\end{table}

Due to  the orders of magnitude lower
densities in compact HII--regions, this effect  is absent there (Gordon \&
Walmsley \cite{GW}), but must be reckoned with in wind sources.
A more exact modelling of the line broadening will be included in the
forthcoming paper (B\'aez--Rubio et al., in prep.) which uses the MORELI
code (B\'aez--Rubio et al. \cite{Moreli}) to model recombination line
and continuum emission.  

\subsection{Stimulated Emission ?}
\label{ss:SE}

\begin{figure}
\centering
\includegraphics[width=0.47\textwidth,angle=0]{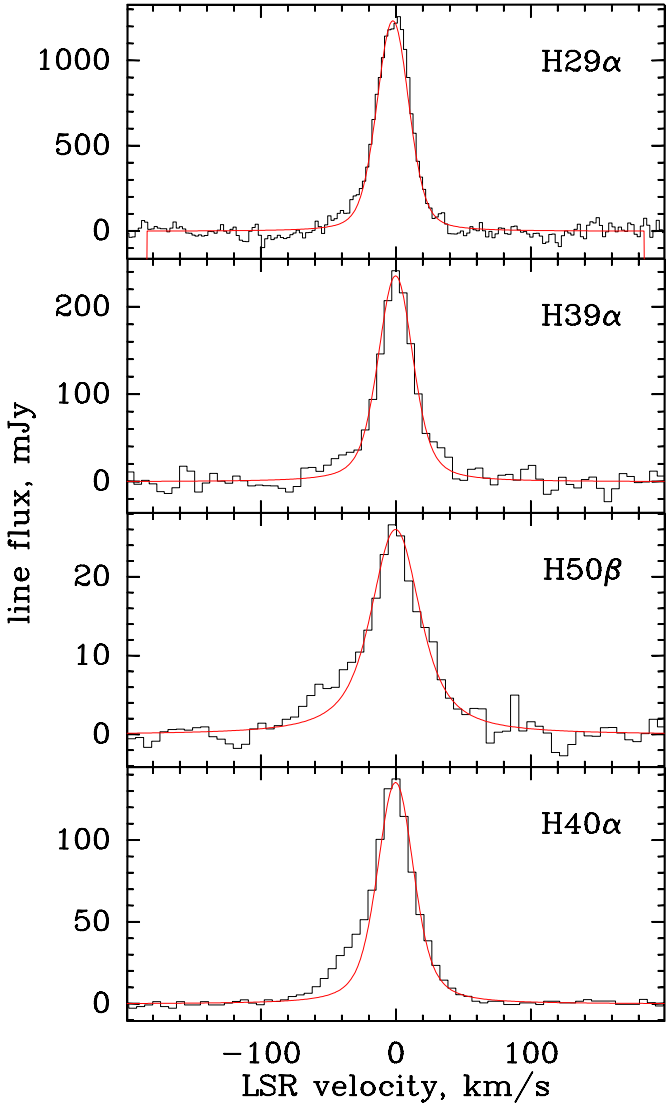}
\caption{Line profiles of the 4 transitions observed with the highest
  signal--to--noise. The underlying continuum  
  emission has been subtracted.  The continuous red lines represent
  Voigt functions (see text). All four transitions display excess
  emission on the blue line wing which is unaccounted for by Doppler
  and impact broadening of the thermal line profiles.
  \label{f:voigt}
 }
\end{figure}

We have argued in sect.~\ref{sss:L/C} on the basis of the integrated
line--to--continuum ratio that the millimeter recombination lines are
emitted close to LTE. Investigation of the line {\it profiles} allows to
carry this discussion a bit further. As pointed out already above, 
all the four best observed lines display emission on their blue wings in
excess of the fitted Voigt profile (Fig.~\ref{f:voigt}). The blue
excess is most pronounced in \Ha{40} and barely detected in \Ha{29}.  

There are two possible causes of the blue excess. Both
scenarios assume that the emission in the wings predominantly comes
from gas in the outflow. In the first scenario, the
lines might be intrinsically symmetric. Excess emission on red wing
may be absorbed by the optically thick continuum core which partially
hides the  receding part of the outflow from the observer.
In the second scenario, the optically thick core enhances the emission
of the approaching outflow cone by stimulation. Such weak 
amplification of line wings has already been seen in \MWC\  where \BR\ et
al. (\cite{Moreli}) called them ``wing humps''.

Detailed modelling of the emission with the MORELI code may help to
decide between the two scenarios, and further observations at wavelengths
longer than 3mm and/or at higher angular resolution may directly show
whether obscuration or stimulated emission is at work. 

%

\section{Discussion}
\label{s:Dis}

The principal characteristic of \Lk\ in the radio/millimeter 
range is its spectral energy distribution which has a constant slope 
through this whole wavelength range. To our knowledge, the only other
source whose slope has been measured and also found to be constant 
over an even larger wavelength range is \MWC.
It is therefore of interest to compare the properties of these two
radio stars (sect.~\ref{ss:mwc}) and to investigate why, despite the overall
similarities between these two stars, their continuum slope is different
(sect.~\ref{ss:wind}), and last but not least, why one star is masing
and the other one not (sect.~\ref{ss:masing}).

\subsection{A non--canonical wind}
\label{ss:wind}

As described in Sect.~\ref{ss:cont}, three independent observations
lead us to the conclusion that the spectral slope $\alpha$ of the continuum is
steeper than expected for an isothermal, isotropical, ionized wind
expanding at constant velocity. The wind in \Lk\ must therefore depart
from the canonical case in at least one of these four properties.

Since the continuum absorption coefficient depends on the electron
temperature as $T_e^{-1.35}$, the emitted flux must also depend on
$T_e$. Gradients of $T_e$ therefore effect $\alpha$.  The dependence
of the emitted flux on temperature is however very weak, $T_e^{0.1}$
(Panagia \& Felli \cite{PF}). It cannot account for the additional
flux observed at 100 GHz due to the
increase of $\alpha$ from the canonical 0.6 to the observed 0.90,
which is a factor of 2.5 compared to 5\,GHz. To explain this factor,
$T_e$ would have to increase by an implausible factor of \mbox{$>\!10^3$} when
approaching the star from $R(5\, \rm GHz)$ to $R(100\, \rm GHz)$. We note
that an opposite $T_e$ gradient, such as due to reheating of the wind at
larger radii, would also affect $\alpha$, but in the wrong sense. We thus
conclude that gradients of $T_e$ cannot explain the observed
steepening of the SED.

Is it possible that the wind is not ionized throughout the flow, as
required by the canonical case ?  If the wind recombines at larger
radii, some of the emission expected at lower frequency is missing,
and the SED is steepening. This is however 
unlikely, since in wind sources virtually all of the
recombinations occur very close to the base of the flow which is
usually optically thick at radio wavelengths (Felli \&
Panagia \cite{FP81}). 
The existence of an HII--region excited by \Lk\ clearly
demonstrates that the star driving the wind
provides more ionizing photons than can be absorbed by the entire flow.

Since departures from spherical symmetry, notably a polar double--cone
geometry,  do not generally change
$\alpha$ (Schmid-Burgk \cite{JSB}), we are thus left with the only
option that the density law of the ionized wind in \Lk\   departs from
$r^{-2}$. The straightforward case where the wind is accelerated leads
to a steeper density dependence and a larger $\alpha$ (see e.g. Panagia \&
Felli \cite{PF}). We consider this case however as very implausible,
since it requires a force acting on the wind uniformly over the very
large range of radii from $R(300\, \rm GHz)$ out to $R(1\, \rm
GHz)$. We are not aware of any force with such characteristics. 

Instead, we propose that the departure from the  $r^{-2}$ law is
related to the presence of a massive circumstellar disk from where the
wind is launched. We envision two possible scenarios. In the first
scenario, the wind originates in a radially narrow ring--like  region 
near the inner edge of the disk, possibly due to photo--evaporation. 
Tuthill et al. (\cite{Tut}) give the radius where
dust evaporates around a star of the luminosity of \Lk\ as $r_C \sim
15$ a.u., in close agreement with the radius where they find the peak
of the dust torus. 
%
The wind  then flows away 
from the disk into cones with opening angles increasing with increasing
height above the disk. The increase of the opening angle is a natural
consequence of the thermal pressure of the wind which causes the plasma
to expand laterally into the neighboring empty space.  The density of
the constant--velocity wind then falls of more rapidly than $r^{-2}$ 
since it expands into  volume segments increasing more rapidly than $r^2$.    
Quantitatively, the increase of the opening angle becomes significant
only in slow winds where \Vexp\ is not much greater than the thermal
speed of the plasma.

The second scenario, not necessarily exclusive of the first one,
postulates that the wind originates from the corona of a massive disk 
whose density falls of exponentially with height above the disk plane,
much like in the photo--evaporative disks modelled by Hollenbach et
al. (\cite{Holl}). We expect that this rapid decrease of the
electron density with distance from the disk generates a steep SED,
even if, to our knowledge, the detailed SED for such
photo--evaporative flows has not yet been modelled.  
The density structure might be complicated even more if the disk
corona contains active regions, like in the solar corona, which may be
instrumental in launching the wind. Although there
are no confirmed reports of a magnetic field in \Lk (Zinnecker \&
Preibisch \cite{ZP}; Osten \& Wolk \cite{OW}; Gibb et
al. \cite{Gibb2010}), a strong magnetic field has 
been discovered in the phenomenologically similar (see next section) \MWC\ 
(Thum \& Morris \cite{field}).     
These authors argue that the magnetic pressure is comparable to the
thermal energy density of the plasma and thus dynamically important.
The field in \MWC\ is claimed to be due to a disk dynamo. As the
parameters under which such dynamos operate are very general (Tout \&
Pringle \cite{TP}), it is
quite possible, that the disk of \Lk\ is also magnetic.

The second scenario gets some support from the consideration that the
gravitational radius $r_g$ (Hollenbach et al. \cite{Holl}), outside of
which plasma at 7000~K is not gravitationally bound anymore by a
15~\Msun\ star is 160 a.u. or 0\farcs22. Since both the millimeter
source (Tab.~\ref{t:cont}) and the near infrared source (Tuthill et
al. \cite{Tut}) are  much smaller,  the wind is obviously launched
from deep inside $r_g$. It then needs an additional force, apart from
thermal pressure, to get started.  A magnetic field can provide this force.

\subsection{\Lk and \MWC: stellar twins ?}
\label{ss:mwc}

The main characteristic of \Lk\ in the radio/millimeter 
range is its SED which has a constant slope through this whole wavelength
range very close to $\alpha= 0.9$.
The SED of \MWC\ 
has been studied over an even larger
wavelength range (Tafoya et al. \cite{Taf}; Krips et al. \cite{Krips}).  
Its slope is also constant ($\alpha = 0.6$), but much closer to the
canonical value. In both cases, the expansion velocity
of the ionized wind is small, $\sim\!60$ \km\ for \MWC\ and 55 \km\ for \Lk,
which sets these disk wind sources apart from the much more numerous class of
mass loosing OB stars whose radiation driven winds reach terminal
velocities of $\simgt1000$ \km\ (Lamers \& Nugis \cite{LN}). It is
curious that the small difference in \Vexp\ between the two sources is
in the same sense as their mass loss rate, a factor 5 higher in \MWC,
and their luminosities, a factor $\simgt5$ higher in \MWC. \\

Similarities between \Lk\ and \MWC\ are strong also at shorter
wavelengths. Both stars have been found by their bright H$\alpha$ (and
other optical) emission lines, they are both massive, have hot
circumstellar disks, and 
their optical spectra are characterized by a complete absence of any
photospheric absorption lines. As a result, the spectral type and
luminosity class of both
stars is inferred only indirectly through their bolometric luminosities
and the presence or absence of ionized helium.  Their evolutionary
state is therefore not well known.  Whereas for \MWC\ 
evidence has accumulated for a post--main--sequence evolutionary state
(Hartmann, Jaffe \& Huchra \cite{HJH}; Hofmann et al. \cite{Hofmann}; 
Gvaramadze \& Menten \cite{GM}),
\Lk\ is likely at or
near the zero age main sequence, as suggested by its close association
with an HII--region and with a cluster of pre--main--sequence stars.
 
Both stars have very broad H$\alpha$ line wings, in both cases
attributed to scattering of line photons in a dense electron gas
(Hartmann et al. \cite{HJH}; Herbig et al. \cite{HAD}). This gas is
thought to be located within a few astronomical units from the star,
possibly also causing the strong veiling of the photospheric
absorption lines. \\

In summary, the two stars display strong phenomenological similarities 
over the electromagnetic spectrum, while one of them is at the
  beginning and the other at the end of their evolution.


\subsection{Millimeter Recombination lines are not masing}
\label{ss:masing}

If \Lk\ bears such a strong similarity to \MWC, why are its millimeter
recombination lines not also masing ? We think that the absence of
masing is due to two adverse circumstances: \1 its disk is smaller
than that of \MWC\ and \2 the disk has an unfavorable orientation.

\JMP\ et al. (\cite{JMP}) have shown that the blue and red \Ha{30} maser spikes
in \MWC\ are located on opposite sides of the circumstellar disk at 29
a.u. from the star. At these positions, the maser spikes are barely
inside the disk of hot dust whose size was measured at
$\lambda3.8\,\mu$m  by Danchi et al. \cite{Danchi} to be 75 a.u.
The \Ha{30} maser in \MWC\ operates at $n_e \simeq$ \dex{3}{7} \ccm\ 
(\JMP\ et al. \cite{MBTW}), 
very close to the density where the net amplification is highest
(Strelnitski et al. \cite{VS1}). At this density, the net line 
absorption coefficient at the line center is $\kappa_L =$
\dex{-1.4}{-12}\,m$^{-1}$ assuming the electron temperature
derived in sect.~\ref{sss:L/C} and no additional line  broadening. 
The unit e--folding gain length is then 
4.7 a.u. Knowing that the maser propagates along tangential lines
above the nearly edge--on disk, it is easy to see that the \MWC\ disk
is large enough to accommodate a few gain lengths leading to a modest
amplification at 1.3\,mm, $\tau\sim\!-3$, as inferred already before (\JMP\ et
al. \cite{MBTW}). 

The disk around \Lk\ is considerably smaller.  
We argued above (sect.~\ref{ss:wind}) that the wind may be launched
from the torus at $r_C\sim\!15$\,a.u. where the emission from hot dust
peaks (Tuthill et al. \cite{Tut}). Assuming that
the plasma density also peaks in this ring, it is reasonable to expect
that maser propagation conditions are optimum there. We then
derive a very modest amplification of $\tau\sim-1$ from simply scaling
with the disk dimensions. But even this very small amplification would only be
achieved if the \Lk\ disk had a near edge--on orientation like
\MWC. This is however not the case. Tuthill et al. infer a likely
inclination angle of $\simlt35$\dgr. This nearly face--on viewing
geometry foreshortens the maser path where it would propagate through the high
densities near the disk surface.  Taking these arguments together, the
conditions for maser amplification in \Lk\ are at least an order of magnitude
worse than in \MWC, and no appreciable amplification results at millimeter
wavelengths.  However, since the hydrogen absorption coefficient
becomes more negative at shorter wavelengths (and higher electron
densities), noticeable amplification may be present at submillimeter
wavelengths.


\section{Conclusions}


Our observations at millimeter wavelengths present the first 
detections of radio recombination lines, including several higher
order transitions,  from the massive young stellar object \Lk. 
The good quality of this data gives a precise value of the stellar
radial velocity, free from complications due to dust obscuration
(Kelly et al. \cite{Kelly}), and
we determine \Vexp = 55 \km\ for the wind expansion velocity and
\dex{1.8}{-6} \Msun\,yr$^{-1}$ for its mass loss rate.
The value of \Vexp\ is much lower than previous claims based on optical/NIR
observations (Cohen et al. \cite{Cohen}; Simon and Cassar \cite{SC};
Hamann and Persson \cite{HP}) 
which were complicated by uncertain amounts of non--Doppler line broadening.  
\Lk\ shares the low \Vexp\ with the recombination line maser source
\MWC, and we argue that the slow and massive ionized wind in \Lk\ is also due to
photo--evaporation from the circumstellar disk as claimed for
\MWC\ (Hollenbach et al. \cite{Holl}; \JMP\ et al. \cite{JMP}). 

From the broadband interferometer spectrum and visibility data we
derive the slope $\alpha = 0.90$ of the continuum spectrum.  This value
which is significantly larger than the canonical value of 0.6 for an
isotropic constant--velocity wind, is compatible with the value derived
from longer wavelength radio data. Such larger than canonical values
of $\alpha$ imply that the density of the wind drops faster than
$r^{-2}$. We propose a scenario where the wind is launched from a
radially narrow part of the circumstellar disk, likely at the radius
where dust sublimates and the disk is flared. As the wind 
expands away from the disk, we postulate that it also expands
laterally resulting in the steeper than $r^{-2}$ drop of the density.  
As more ionized winds from massive young stellar objects become
detectable, we expect that their radio/millimeter continuum spectral
slopes will often be steeper than canonical, particularly in those sources
which are in the evolutionary stage where they disperse the gaseous
component of their accretion disks. Steepened SEDs have already been seen in a
number of radio stars (Altenhoff et al. \cite{ATW}) and a few highly
compact HII--regions (Franco et al. \cite{Franco}).

Our data of the higher order recombination lines, the first such
observations in a stellar wind source apart from \MWC, display a
novel phenomenon. All ratios of close frequency pairs 
(Fig.~\ref{f:ratios}) are significantly lower than 
observed in compact HII--regions. We
show that this trend most likely results from the circumstance that
the higher order lines, emitted from levels of higher principal
quantum numbers $n$ than the corresponding $\alpha$--transition, are
affected by pressure broadening in the high densities of the
ionized wind. We plan to study this novel phenomenon more
quantitatively in a forthcoming paper using the modelling code MORELI
(Baez--Rubio et al. \cite{Lk}).

Despite the strong phenomenological similarities between \Lk\ and \MWC,
the recombination lines in \Lk\ are emitted near LTE. 
The blue line wings may however be weakly amplified, possibly like 
the ``wing humps'' in \MWC. The absence of strong {\it disk} masers
at millimeter wavelengths is quantitatively explained by the smaller
size of the \Lk\ disk and its unfavorable
orientation. Notwithstanding, noticeable amplification may be
detectable at submillimeter and shorter wavelengths if the required
high electron densities ($\sim$\dex{}{8} \ccm) are present in this source.

\begin{acknowledgements}
 We are grateful to the IRAM Director, P. Cox, for granting 
 time on the Plateau de Bure interferometer 
 in the context of commissioning of the broad band correlator
 WIDEX. We thank Malcolm Walmsley for valuable comments on impact
 broadening and an anonymous referee for insightful remarks.
\end{acknowledgements}



\begin{thebibliography}{}
 \bibitem[1981]{ASW}
     Altenhoff, W.J., Strittmatter, P.A., \& Wendker, H.J. 1981, \aap\vol{93} 48
 \bibitem[1994]{ATW} 
     Altenhoff, W.J., Thum, C., \& Wendker, H.J. 1994, \aap\vol{281} 161
 \bibitem[2013]{Moreli}
     B\'aez--Rubio, A., \JMP, Thum, C., \& Planesas, P.   2013, 
     \aap\vol{553} A45
 \bibitem[2013]{Lk}
     B\'aez--Rubio et al. 2013, in preparation.
 \bibitem[1990]{Barsony}
     Barsony, M., Scoville, M.Z., Schloembert, J.M., \& Claussen,
     M.J. 1990, \apj\vol{362} 674
 \bibitem[1988]{BW}
     Becker, R. H., \& White, R. L. 1988, \apj\vol{324} 893
 \bibitem[1972]{BS}
     Brocklehurst M. \& Seaton, M.J. 1972, \mnras\vol{157} 179---210
 \bibitem[2012]{Matt} 
     Carter, M., Lazareff, B., Maier, D., Chenu, J.-Y. et al. 2012, 
     \aap\vol{538} 89
 \bibitem[1980]{Cohen}
     Cohen, M. 1980, \mnras\vol{190} 865 
 \bibitem[1982]{CBS}
     Cohen, M., Bieging, J. H., Schwartz, P. R. 1982, \apj\vol{253} 707C\	
 \bibitem[2001]{Danchi}
     Danchi, W.C., Tuthill, P.G., \& Monnier, J.D. 2001, \apj\vol{562} 440
 \bibitem[1986]{DR}
     Dewdney, P. E., \& Roger, R. S. 1986, \apj\vol{307} 275
 \bibitem[1981]{FP81}
     Felli, M. \& Panagia, N. 1981, \aap\vol{102} 424
 \bibitem[2000]{Franco} 
     Franco, J., Kurtz, S., Hofner, P., et al.\ 2000, \apjl\vol{542} L143 
 \bibitem[2012]{GM}
     Gvaramadze, V.V., \& Menten, K.M.  2012, \aap\vol{541} 7
 \bibitem[2010]{Gibb2010}
     Gibb, E. L., Brittain, S. D., Rettig, T. W., Troutman, M., Simon,
     Theodore,  Kulesa, C. 2010, \apj\vol{715}, 757
 \bibitem[2007]{GH}
     Gibb, A.G., Hoare, M.G. 2007, \mnras\vol{380} 246
 \bibitem[1990]{GW}
     Gordon, M.A. \& Walmsley, C.M. 1990, \apj\vol{365} 606
 \bibitem[1989]{HP}
     Hamann, F. \& Persson, S.E. 1989, \apjs\vol{71} 931
 \bibitem[1979]{Harvey}
     Harvey, P.M., Thronson, H.A., Gatley, I. 1979, \apj\vol{231} 115
 \bibitem[1980]{HJH}
     Hartmann,L., Jaffe, D., Huchra, J.P. 1980, \apj\vol{239} 905
 \bibitem[1971]{Herbig}
     Herbig, G.H. 1971 \apj\vol{169} 537
 \bibitem[2004]{HAD}
     Herbig, G.H., Andrews, S.M., \& Dahm, S.E. 2004, \aj\vol{128} 1233
 \bibitem[2002]{Hofmann} 
     Hofmann, K.-H., Balega, Y., Ikhsanov, N.~R., Miroshnichenko,
     A.~S.,  \& Weigelt, G.\ 2002, \aap\vol{395}  891 
 \bibitem[1994]{Holl}
     Hollenbach, D., Johnstone, D., Lizano, L. \& Shu, F. 1994,
     \apj\vol{248} 654
 \bibitem[1994]{Kelly}
     Kelly, D.M., Rieke, G.H., Campbell, M. 1994, \apj\vol{425} 231
 \bibitem[1976]{Knapp}
     Knapp, G. R., Kuiper, T. B.H., Knapp, S. L., \& Brown, R. L. 1976, 
     \apj\vol{206} 443
 \bibitem[2013]{Krips}
     Krips, M., Neri, R., Moreno, R., Thum, C. et al.  2013, in preparation
 \bibitem[2002]{LN}
     Lamers, H. J. G. L. M. \& Nugis, L. 2002, \aap\vol{395} L1 
 \bibitem[1989]{MBTW}
     Mart\'\i n--Pintado, J., Bachiller, R., Thum, C. \& Walmsley, C.~M. 1989,
     \aap\vol{215}  L13  
 \bibitem[2011]{JMP}                 
     Mart\'\i n--Pintado, Thum, C., Planesas, P., \& Baez--Rubio, A. 2011,
     \aap\vol{530}  L15 
 \bibitem[1990]{Mitchell}
     Mitchell, G. F., Maillard, J.-P., Allen, M., Beer, R., \&
     Belcourt, K. 1990, \apj\vol{363} 554
 \bibitem[1967]{MH}
     Mezger, P. G., \& H\"oglund, B. 1967, \apj\vol{147} 490
 \bibitem[1975]{Olnon}
     Olnon, F.M. 1975, \aap\vol{39} 217
 \bibitem[2009]{OW}
     Osten, R.A., Wolk, S.J. 2009, \apj\vol{691} 1128
 \bibitem[1973]{Nino}
     Panagia, N.  1973, \aj\vol{78} 929
 \bibitem[1975]{PF}
     Panagia, N. \& Felli, M. 1975, \aap\vol{39} 1
 \bibitem[1999]{Pirogov}
     Pirogov, L. 1999, \aap\vol{348} 600
 \bibitem[1992]{RRLreview}
     Roelfsema, P. R. \& Goss, W. M. 1992, \anrev\vol{4} 161
 \bibitem[1982]{Luis}
     Rodr{\'\i}guez, L.F. 1982, \rmex\vol{5} 179
 \bibitem[2009]{RZH}
     Rodr{\'\i}guez, L.F., Zapata, L.A. \& Ho, P.T.P 2009,
     \apj\vol{692} 162-–167
 \bibitem[2011]{Sandell}
     Sandell, G., Weintraub, D. A., Hamidouche, M. 2011, \apj\vol{727} 26
 \bibitem[1982]{JSB}
     Schmid-Burgk, J. 1982, \aap\vol{108} 169
 \bibitem[2011]{Sewilo}
     Sewilo, M., Churchwell, E., Kurtz, S., Goss, W. M., Hofner,
     P. 2011,  \apjs\vol{194} 44
 \bibitem[1984]{SC}
     Simon, M.  \& Cassar, L. 1984, \apj\vol{283} 179
 \bibitem[1996]{VS1}
     Strelnitski, V., Ponomarev, V., Smith, H.A. 1996 \apj\vol{470} 1118
 \bibitem[2004]{Taf}
     Tafoya, D, G\'omez, Y. \& Rodr{\'\i}guez, L.F. 2004,
     \apj\vol{610} 827
 \bibitem[1976]{Thompson}
     Thompson, R.I., Erickson, E.F., Witteborm, F.C., \& Strecker, D.W. 1976, 
     \apj\vol{210} L13
 \bibitem[1998]{laser} 
     Thum, C., Mart\'\i n--Pintado, J., Quirrenbach, A. \& Matthews,
     H.E.  1998,      \aap\vol{333} \rm{L}63 
\bibitem[1995]{beta}
     Thum, C.,  Strelnitski, V.S., Matthews, H.E., Mart\'\i n--Pintado, J., 
     \& Smith, H.A. 1995, \aap\vol{300} 843 -- 850
 \bibitem[1999]{field}
     Thum, C., Morris, D. 1999, \aap\vol{344} 923 -- 929 
 \bibitem[1992]{TP}
     Tout, C.A. \& Pringle, J.E. 1992, \mnras\vol{259} 604
 \bibitem[2002]{Tut}
     Tuthill, P.G., Monnier, J.D., Danchi, W.C., Hale, D.D., Townes,
     C.H. 2002, \apj\vol{577} 826
 \bibitem[1990]{Malcolm}
     Walmsley, C.M. 1990, \aas\vol{82} 201
 \bibitem[1994]{ZP}
     Zinnecker, H. \& Preibisch, Th. 1994, \aap\vol{292} 152
\end{thebibliography}
\end{document}